# Graviton: interchain swaps and wrapped tokens liquidity incentivisation solution

Aleksei Pupyshev, Ilya Sapranidi, Elshan Dzhafarov, Shamil Khalilov, Ilya Teterin

*VenLab team*

aleksei@venlab.dev

## Abstract

This paper discusses the issues with liquidity that inhibit adoption of so-called wrapped tokens, i.e. digital assets issued in one blockchain ecosystem (origin) with representation in other blockchain networks (destination), and an incentive model and a governance mechanism for solving these issues are suggested. The proposed liquidity model called Graviton can be implemented both within the framework of a single destination chain, or as a blockchain-agnostic solution combining various blockchain platforms together and providing liquidity to wrapped tokens in each of them. This model does not depend on how cross-chain transfer gateways are implemented, and can work with both centralized gates and bridges, or decentralized trustless gateways, as well as gateways based on oracle networks and threshold signatures.

# Introduction

Today, there exist a large number of popular blockchain networks with native tokens or non-native tokens of various standards. Each blockchain platform seems to have its own advantages and disadvantages in technical terms, such as the size of fees, time scales for block generation and transaction finalization, the availability of decentralized applications and tools for communicating with the blockchain network (wallets, browsers, browser plugins), as well as the community, partners and market makers.

Recently, decentralized exchanges based on automatic market-making (AMM) models through liquidity pools have been gaining popularity. The prices on such exchanges are formed in a trustless manner and depend only on the ratio of the amounts of one token paired with another. When the balances change, the price is automatically adjusted so that equivalence is preserved (for example, the product of amounts can be kept constant, as in Uniswap [1]). There are multiple ways to establish equivalence formulas used in these AMM models [2], and this creates many arbitrage opportunities and attracts algorithmic traders, thereby increasing the efficiency of the digital asset market as a whole.

There are a number of incentive models for shareholders in liquidity pools to hold pairs of assets on a smart contract. Most often, this is implemented as a fee subtracted from trading operations, which is added together and distributed to the shareholders of a liquidity pool. However, another model, namely a reward token, has recently begun to gain popularity, which can simultaneously play the role of a governance token, capable of influencing the development of a protocol.

The presence of a large number of diverse platforms, on the one hand, brings diversity to the industry of public distributed ledgers and promotes competition between ecosystems. On the other hand, it creates barriers to the mass adoption of decentralized applications and digital assets, since each platform has its own separate interfaces, explorers or browser applications, in addition to having a different level of support by centralized services, exchanges and wallets. Interoperability between different blockchain platforms is an important goal, and its achievement can create and sustain a major network effect on the entire crypto industry as a whole.

The first major practical example of interoperability is the capacity to execute cross-chain transfers of digital assets. This opens up a number of opportunities, as different groups of users, traders, or algorithmic trading systems are becoming active in different blockchain ecosystems. Differences in liquidity, block generation speed, trading activity, and volumes, create arbitrage opportunities that may be of interest for traders.

In this paper, arbitrage opportunities are understood in the broadest sense, including the emergence of any market inefficiency, such as the difference in prices on centralized or decentralized exchanges, the difference in interest rates on lending services, and the difference in staking profitability in various staking or liquidity pools.

The presence of different blockchain versions of the same digital asset can lead to a more widespread usage of it in decentralized applications of various blockchain platforms, where their audiences that consist of active agents (traders, bots, users) can overlap only slightly.

Thus, interchain versions of tokens, in addition to providing convenience for users, also provide an opportunity to earn on market inefficiencies that arise between platforms. Such opportunities create positive feedback in the systems, leading to increased liquidity and a larger number of transactions, and ultimately to the network effect for all integrated platforms that increases adoption.

Usually, a digital asset has a blockchain that it was first issued on or where it is used as the main utility token. This blockchain would constitute the origin for the token, whereas tokens that are issued on a non-origin blockchain are called **wrapped** tokens in a destination blockchain.

## Problems

Despite the fact that wrapped tokens on their own represent a solution to the interoperability problem, there are a number of issues that limit their widespread use:

1. Technical limitations and risks of cross-chain gateways implementation
2. Low availability and popularity of services that can provide seamless cross-chain transfers and that support operations with wrapped tokens in the destination blockchains
3. Lack of liquidity of wrapped tokens in destination blockchains

The first issue can be solved by implementing trustless bridges, as done in the SuSy protocol, Rainbow, or ZK & MPC solutions [4]. The second issue has a strong dependency on the third, which conversely depends on the second, however, we suggest that the primary issue that needs to be solved is the lack of liquidity of wrapped tokens in destination chains. This article is fully devoted to presenting a way to solve the key challenge associated with liquidity.

The term liquidity is presented here in the broadest sense: as the ability to purchase a token or exchange it for another digital asset with minimal capital losses and for the maximum possible transaction volume. The latter assumes both the usage of centralized services such as exchanges, OTC and lending [3] platforms with minimal spread volumes (CeFi-liquidity), and the presence of a wide range of decentralized AMM DEXes, orderbook-based DEXes, OTC or lending decentralized applications, with large amount of funds deposited in them (DeFi-liquidity).

## Proposed Solution

Our proposed solution can only work in the context of DeFi-liquidity due to its decentralized nature, automation through smart contracts and composability between different decentralized applications. In particular, only mechanisms for working with the so-called AMM DEXes will be considered. However, the proposed protocol can potentially be extended to other types of DeFi dApps (lending protocols, vaults, OTCs, hybrid DEXes, order-book DEXes and etc).

AMM DEXes (Automated Market Making Decentralized Exchanges) operate on the basis of varying dynamics of token amounts in liquidity pools. In order to replenish a liquidity pool, it is necessary to lock both tokens that form the pair at the current pool price on the pool's smart contract. The more liquidity there is in the pool, the less the slippage is: that is, the losses that can occur when buying one token for another, in comparison with the fair market price.

The essence of the Graviton protocol solution is to create incentives for both cross-chain transfer providers and shareholders of wrapped tokens' liquidity pools for the most popular and liquid tokens of the destination chain.

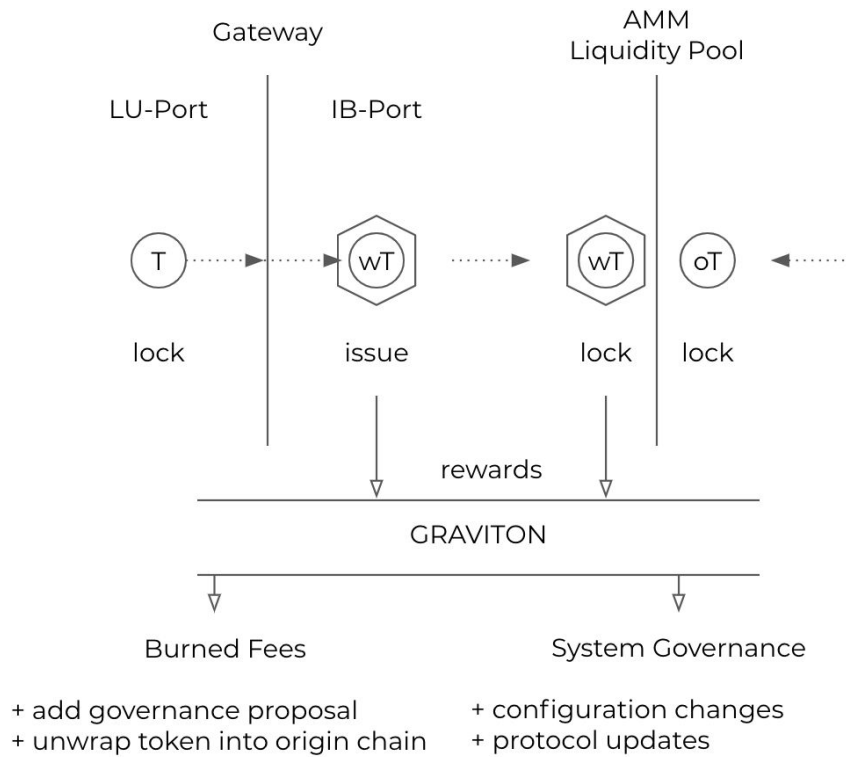

Fig 1. Scheme of incentivisation of the Graviton protocol based on Reward, Governance & Utility (**RGU**) token

Figure 1 illustrates a scheme for the incentivization protocol: a token (**T**) in the origin blockchain is sent to a gateway, which consists of a receiving LU-Port (lock & unlock) [4] on which the token is locked, as well as an IB-Port (issue & burn) where the wrapped token (**wT**) is issued in the destination blockchain. To issue a wrapped token, it needs to be blocked on the origin blockchain and issued on the destination blockchain.

In order for users to gain access to the wrapped token, it must be listed on decentralized exchanges. To place liquidity on AMM exchanges in liquidity pools, in addition to listing a wrapped token, it is required to list a different origin token that is widely used in the destination chain (**oT**) as a pair. The best candidate for this role is either the native token of the destination chain or the most popular and liquid stablecoin.

The scheme described above is widely known and in itself is not sufficient for a fully practical solution to the problem of liquidity of wrapped tokens in destination chains. However, when introducing a new digital asset that simultaneously combines reward, governance and utility functionality (**RGU token**), it becomes possible to solve the issues with liquidity. Depending on the specific implementation, the presence of a developed inter-chain communication infrastructure (for example, the Gravity protocol [5]) and depending on the pursued goals, the **RGU token** can be a token of one origin chain only, or be an interoperable token, conforming to the dualism of origin-destination for any blockchain integrated into the system.

The most important aspect of the implementation is that the reward should be calculated based on two fundamental factors, such as the **amount** of wrapped tokens locked in liquidity pools, as well as the **time** when it was locked in the pool. Rewards are accrued both by gateway providers themselves and by the wT/oT pair pool shareholders on an AMM exchange. The distribution of rewards to gateway providers also depends only on the amount **wT** in the pools, in order to prevent possible "excessive farming" attacks from the cross-chain transfer gateway providers themselves.

## Utility

By itself, an **RGU token** has no value as long as it has no fundamental utility functions in the system. Moreover, when generating rewards alone, its supply will increase, making it even less valuable, at the same time disrupting the very scheme of incentivizing liquidity providers. Endowing a token with a payment functionality within the system, as well as allowing to burn it after a payment is made, is a fundamental property that gives **RGU token** value.

Burning the token received from the commissions for the reverse swap (un-wrapping) leads to incentivizing the wrapped token to be held in the destination chain. Moreover, this helps to maintain its liquidity and at the same time allows for a withdrawal to the target chain, charging a user fee and increasing the value of the token by stabilizing its total supply.

When considering the Graviton system as a network of entities with different interests and the goal to preserve the decentralized nature of its technical development, protocol updates, and tuning of configuration parameters, a governance mechanism is required, implemented through a mechanism of proposals from the community of token holders. The initiation of proposals should also be paid for in the **RGU token**. Such a mechanism, in addition to increasing the utility of the token, also protects against spam activity on the part of protocol detractors.

## Governance

A decentralized governance model would allow the system to develop and update flexibly, adjusting to the changing market conditions and technologies, thereby giving additional incentive to accumulate and own the **RGU token.** The possible governance functions can be categorized through various types of proposals (GIP: Graviton improvement proposals):

- Voting for updating system parameters, such as reward distribution formulas and various weights for liquidity pools, origin of tokens (oT), blockchain networks, as well as ratios of rewards with regards transfer and liquidity providers. The regular general emission is also determined by a separate formula, the parameters of which are subject to tuning and adaptation to market conditions;

- Voting for extending the system through the integration of new wrapped tokens, new origin tokens, new liquidity pools, as well as support for new blockchain networks;

- Voting for new functionality of the protocol itself: for instance, the introduction of lending, OTC or framing services as new decentralized tools for maintaining the liquidity of wrapped tokens. Integration of various services and cross-chain communication protocols can also be a part of such updates.

Voting mechanics are not crucial to the protocol functionality and can be implemented in any conventional way.

## Conclusion

The Graviton protocol, while not presenting a universal solution to the problem of inter-chain composability, in combination with other lower-level protocols (oracles and cross-chain gateways) allows for the creation of an integral system aimed at stimulating the expansion of the concept of wrapped tokens. Thus, it represents an important step on the way to achieve full inter-chain composability.